\documentclass[pra,aps,groupaddress,twocolumn]{revtex4}
\usepackage{epsfig,amsmath,graphicx,amssymb,overpic}
\usepackage{bm}

\def\be{\begin{equation}}
\def\ee{\end{equation}}
\def\bee{\begin{eqnarray}}
\def\ene{\end{eqnarray}}
\def\bes{\begin{subequations}}
\def\ees{\end{subequations}}

\begin{document}

\title{Attraction centers and $\mathcal{PT}$-symmetric delta-functional
dipoles \\
in critical and supercritical self-focusing media}
\author{Li Wang$^{1,2}$}
\author{Boris A. Malomed$^{3}$}
\author{Zhenya Yan$^{1,2,}$}
\email{zyyan@mmrc.iss.ac.cn}
\affiliation{\vspace{0.1in} $^1$Key Lab of Mathematics Mechanization, Academy of
Mathematics and Systems Science, Chinese Academy of Sciences, Beijing
100190, China\\
$^2$ School of Mathematical Sciences, University of Chinese Academy of
Sciences, Beijing 100049, China\\
$^{3}$Department of Physical Electronics, School of Electrical Engineering,
Faculty of Engineering, and Center for Light-Matter Interaction, Tel Aviv
University, Tel Aviv, 59978, Israel}

\begin{abstract}
%\vspace{0.1in}
We introduce a model based on the one-dimensional nonlinear Schr\"{o}dinger
equation (NLSE)\ with the critical (quintic) or supercritical self-focusing
nonlinearity. We demonstrate that a family of solitons, which are unstable
in this setting against the critical or supercritical collapse, is
stabilized by pinning to an attractive defect, that may also include a
parity-time ($\mathcal{PT}$)-symmetric gain-loss component. The model can be
realized as a planar waveguide in nonlinear optics, and in a
super-Tonks-Girardeau bosonic gas. For the attractive defect with the
delta-functional profile, a full family of the pinned solitons is found in
an exact analytical form. In the absence of the gain-loss term, the
solitons' stability is investigated in an analytical form too, by means of
the Vakhitov-Kolokolov criterion; in the presence of the $\mathcal{PT}$%
-balanced gain and loss, the stability is explored by means of numerical
methods. In particular, the entire family of pinned solitons is stable in
the quintic (critical) medium if the gain-loss term is absent. A stability
region for the pinned solitons persists in the model with an arbitrarily
high power of the self-focusing nonlinearity. A weak gain-loss component
gives rise to intricate alternations of stability and instability in the
system's parameter plane. Those solitons which are unstable under the action
of the supercritical self-attraction are destroyed by the collapse. On the
other hand, if the self-attraction-driven instability is weak and the
gain-loss term is present, unstable solitons spontaneously transform into
localized breathers, while the collapse does not occur. The same outcome may
be caused by a combination of the critical nonlinearity with the gain and
loss. Instability of the solitons is also possible when the $\mathcal{PT}$%
-symmetric gain-loss term is added to the subcritical nonlinearity. The
system with self-repulsive nonlinearity is briefly considered too, producing
completely stable families of pinned localized states. \vspace{0.2in}
\end{abstract}

\maketitle

\baselineskip=13pt

\section{Introduction and the model}

It is well established that non-Hermitian Hamiltonians, subject to the
constraint of the parity-time ($\mathcal{PT}$) symmetry, may satisfy the
fundamental condition of the reality of energy spectra, hence they may be
physically relevant objects~\cite{bender1}-\cite{ptqm}. A single-particle $%
\mathcal{PT}$-symmetric Hamiltonian usually contains a complex potential,
\begin{equation}
U(\mathbf{r})\equiv V(\mathbf{r})+iW(\mathbf{r}),  \label{U}
\end{equation}%
whose real $V(\mathbf{r})$ and imaginary $W(\mathbf{r})$ parts are,
respectively, even and odd functions of coordinates, i.e.,~
\begin{equation}
U^{\ast }(\mathbf{r})=U(-\mathbf{r}),  \label{minus}
\end{equation}%
where the asterisk stands for the complex conjugate. Usually, the energy
spectrum generated by the $\mathcal{PT}$-symmetric potential remains real
below a threshold value of the amplitude of its imaginary part, $W(\mathbf{r}%
)$ in Eq. (\ref{U}), above which the $\mathcal{PT}$ symmetry breaks down
\cite{breaking,Nixon}. Nevertheless, examples of systems with unbreakable $%
\mathcal{PT}$ symmetry are known too \cite{unbreakable}. In fact, the
linearized version of the model introduced in the present work also avoids
the breakdown, see Eqs. (\ref{linear}) and (\ref{linear2}) below.

While the $\mathcal{PT}$ symmetry was not experimentally implemented in
quantum systems (and it was argued that it does not hold in the framework of
the quantum field theory \cite{Szameit}), the possibility to realize the $%
\mathcal{PT}$ symmetry in classical photonic media with mutually balanced
spatially separated gain and loss elements was elaborated both theoretically
\cite{theo1}-\cite{yan16} and experimentally \cite{exp1}-\cite{exci3}. In
addition to that, the same concept can be realized in optomechanics \cite{om}%
, acoustics~\cite{acoustics, electronics}, magnetism \cite{magnetism}, and
Bose-Einstein condensates \cite{Cartarius1,Cartarius2,added}.

Being a linear feature, the $\mathcal{PT}$ symmetry often occurs in
combination with the Kerr nonlinearity of optical media in which it is
implemented. The respective model amounts to the nonlinear Schr\"{o}dinger
equation (NLSE) with a complex potential which is subject to condition (\ref%
{minus}). The NLSE gives rise to $\mathcal{PT}$-symmetric solitons, which
were a subject of intensive theoretical work, see, e.g., original papers
\cite{soliton}, \cite{Konotop}-\cite{SciRep}, \cite{yan15,yan16,added, mixed}%
, and reviews \cite{review1,review2}. The existence of $\mathcal{PT}$%
-symmetric solitons was also experimentally demonstrated in optical lattices
\cite{exp7}. Although $\mathcal{PT}$-symmetric media are, as a matter of
fact, dissipative ones, solitons in these media appear in continuous
families, which is typical for conservative models \cite{families,zezyu}. It
is relevant to stress that $\mathcal{PT}$-symmetric solitons lose their
stability at a critical value of the strength of the gain-loss terms which
is smaller than the above-mentioned threshold value, above which the $%
\mathcal{PT}$ symmetry breaks down in the given system. Between these
values, the solitons still exist, but they are unstable \cite%
{Radik,Konotop,Yang,Sukho}.

A specific form of the one-dimensional (1D) $\mathcal{PT}$-symmetric
potential (\ref{U}) is represented by a delta-functional \textit{dipole},
namely,%
\begin{equation}
U(x)=-[\varepsilon \delta (x)+i\gamma \delta ^{\prime }(x)]  \label{delta}
\end{equation}%
[$\delta ^{\prime }(x)$ stands for the first-order derivative of the
delta-function], which was considered in various contexts (including,
naturally, a regularized version of the delta-function) \cite{Cartarius1}-%
\cite{Kuriakose}. In particular, an advantage offered by potential (\ref%
{delta}) is that it allows one to construct exact solutions for solitons
pinned to the $\mathcal{PT}$-symmetric defect embedded in a spatially
uniform nonlinear medium \cite{Cartarius1, Thaw}, although the stability of
such states, which is a crucially important issue in the context of $%
\mathcal{PT}$ symmetry, was addressed by means of numerical methods.
Actually, models of this type were previously elaborated only for embedding
media with the cubic self-focusing or defocusing nonlinearity. While it
indeed represents the most common type of the self-interaction in photonics
and other settings which admit the realization of the $\mathcal{PT}$
symmetry, higher-order nonlinearities also occur in optics. In particular,
it was experimentally demonstrated that combinations of cubic, quintic, and
septimal terms in the optical response of colloidal waveguides can be
efficiently engineered by adjusting the size and density of metallic
nanoparticles in the colloidal suspension \cite{Cid1,Cid2}. This technique
makes it possible to create an optical medium with a nearly pure quintic or
septimal nonlinearity of either sign. On the other hand, a 1D NLSE with the
quintic self-attraction is an approximate model of the super-Tonks-Girardeau
state, i.e., a quantum gas of strongly interacting bosons in a highly
excited state \cite{superTG}. The attraction center in the bosonic gas can
be implemented, as usual, by means of a tightly focused red-detuned laser
beam. The consideration of these possibilities is relevant, in particular,
because the quintic self-focusing in 1D gives rise to the \textit{critical
collapse} and one-dimensional solitons of the \textit{Townes type} \cite%
{1D-Townes}, suggesting one to look for possibilities to stabilize such
solitons, which are unstable in uniform media \cite{Gaeta,Fibich}.

The objective of the present work is to consider solitons pinned to the $%
\mathcal{PT}$-symmetric defect (\ref{delta}), embedded in the 1D uniform
medium with general self-focusing nonlinearity, of power $2\sigma +1$ with $%
\sigma >0$ ($\sigma =1/2,\,1$, and $2$ correspond, respectively, to the
quadratic, cubic, and critical quintic self-focusing nonlinearities), the
respective scaled NLSE for the wave amplitude $\psi $ taking the form of%
\begin{equation}
i\psi _{z}=-\frac{1}{2}\,\psi _{xx}-\left\vert \psi \right\vert ^{2\sigma
}\!\!\psi -[\varepsilon \delta (x)+i\gamma \delta ^{\prime }(x)]\psi,
\label{NLS}
\end{equation}%
where $z$ and $x$ are the propagation distance and transverse coordinate, in
terms of the underlying optical planar waveguide, the nonlinearity
coefficient is normalized to be $+1$ (which implies self-focusing), $\psi
_{xx}$ represents the paraxial diffraction, and $\varepsilon >0$ accounts
for the attraction strength of the defect (it can be realized by means of
the available technique \cite{Kip}, implanting resonant dopants inducing a
high refractive index doping in a narrow stripe of the waveguide along the $%
z $ axis), while its gain-loss component is accounted for by coefficient $%
\gamma \geq 0$. The latter ingredient of the model can be implemented, also
by means of the doping technique, as parallel narrow amplifying and
absorbing stripes, separated by a small distance, the respective transverse
sizes being on the order of one or few wavelengths of the light beam, while
the use of the paraxial propagation equation (\ref{NLS}) implies that the
transverse size of the beam is much larger than the wavelength, therefore
the $\delta $-functions and its derivative offer a natural approximation in
this case. The above-mentioned model of the super-Tonks-Girardeau gas may be
approximated by Eq.~(\ref{NLS}) with $z$ replaced by scaled time, $t$, the
quintic self-attraction ($\sigma =2$), and $\gamma =0$. The remaining
scaling invariance of Eq.~(\ref{NLS}) allows one to fix $\varepsilon \equiv
1 $, replacing the original variables by
\begin{equation}
\tilde{\psi}=\varepsilon ^{1/\sigma }\psi,\quad ~\tilde{z}=\varepsilon
^{2}z,\quad \tilde{x}\equiv \varepsilon x.  \label{tilde}
\end{equation}%
This rescaling does not affect $\gamma $.

The analysis reported below includes the case of the attraction center in
the conservative medium, i.e., $\gamma =0$, which was previously studied
only for the linear and cubic embedding media ($\sigma =0$ or $1$). The
present model with $\gamma =0$ and higher-order nonlinearity, $\sigma >1$,
including the above-mentioned critical (quintic) case, $\sigma =2$, and
\textit{supercritical} one, $\sigma >2$, makes it possible to construct a
full family of exact solutions for solitons pinned to the attractive defect,
and predict their stability in an analytical form, by means of the
Vakhitov-Kolokolov (VK) criterion, which applies to a broad class of
conservative models with self-attraction \cite%
{VKC,perturbation,Berge,Yang-book,Fibich} (its generalization for solitons
in self-repulsive media, in the form of the \textit{anti-VK criterion}, is
known too~\cite{anti-VK}). Analytical solutions in the form of a family of
pinned solitons are also obtained for the $\mathcal{PT}$-symmetric defect,
with $\gamma >0$, although their stability is investigated by means of a
numerically implemented method, with the delta-function replaced, as usual,
by its regularized version, see Eq. (\ref{approx}) below.

The rest of the paper is arranged as follows. The general theory and
analytical results, including the exact solutions for the delta-functional
defect, application of the VK criterion for the stability analysis, and also
a brief summary of results for the model with the self-repulsive
nonlinearity, are presented in Section II. Numerical findings, including
solution of the stability problem through the computation of eigenvalues for
small perturbations, and simulations of the evolution of unstable solitons,
are reported in Section III. The paper is concluded by Section IV.

\section{General theory and analytical results}

\subsection{Stationary states}

Stationary localized states with propagation constant $k>0$ are looked for
as solutions to Eq. (\ref{NLS}) in the form of
\begin{equation}
\psi (x,z)=\mathrm{e}^{ikz}\phi (x),  \label{uU}
\end{equation}%
with complex function $\phi (x)$ satisfying the ordinary differential
equation with the singular $\mathcal{PT}$-symmetric potential,%
\begin{equation}
-k\phi +\frac{1}{2}\frac{d^{2}\phi }{dx^{2}}+|\phi |^{2\sigma }\phi +[\delta
(x)+i\gamma \delta ^{\prime }(x)]\phi =0  \label{ODE}
\end{equation}%
[recall $\varepsilon \equiv 1$ is fixed in Eq. (\ref{NLS}) by the rescaling (%
\ref{tilde})]. The complex function $\phi (x)$ in Eq.~(\ref{ODE}) can be
rewritten as~\cite{Zhenya}
\begin{equation*}
\phi (x)=\varphi (x)\exp \left[ i\int_{-\infty }^{x}v(x^{\prime })dx^{\prime
}\right]
\end{equation*}%
with real amplitude $\varphi (x)$ and local wavenumber $v(x)$ satisfying
coupled nonlinear ordinary differential equations:
\begin{equation}
\begin{array}{l}
\varphi _{xx}^{\prime \prime }+2\varphi ^{2\sigma +1}+\left[ 2\delta
(x)-v^{2}-2k\right] \varphi ,\vspace{0.1in} \\
(\varphi ^{2}v)_{x}^{\prime }=-2\gamma \delta ^{\prime }\varphi ^{2}.%
\end{array}%
\end{equation}%
Stability of the stationary states is explored by considering their
perturbed version,
\begin{equation}
\psi (x,z)=\left\{ \phi (x)+\eta \left[ F(x)\mathrm{e}^{i\omega z}+G^{\ast
}(x)\mathrm{e}^{-i\omega ^{\ast }z}\right] \right\} \mathrm{e}^{ikz},
\label{pert}
\end{equation}%
where $\eta $ is an amplitude of the infinitesimal perturbation with a
(generally, complex) eigenvalue, $\omega $, and components $F(x)$ and $G(x)$%
, which satisfy the matrix equation, produced by the substitution of ansatz (%
\ref{pert}) and linearization:
\begin{equation}
\left(
\begin{array}{cc}
\hat{L}_{1} & \hat{L}_{2}\vspace{0.05in} \\
-\hat{L}_{2}^{\ast } & -\hat{L}_{1}^{\ast }%
\end{array}%
\right) \left(
\begin{array}{c}
F(x)\vspace{0.05in} \\
G(x)%
\end{array}%
\right) =\omega \left(
\begin{array}{c}
F(x)\vspace{0.05in} \\
G(x)%
\end{array}%
\right) .  \label{lin}
\end{equation}%
Here $\ast $ stands for the complex conjugate, and the constituent operators
are
\begin{eqnarray}
\hat{L}_{1} &=&\frac{1}{2}\partial _{x}^{2}+\delta (x)+i\gamma \delta
^{\prime }(x)+(\sigma +1)\left\vert \phi \right\vert ^{2\sigma }-k,  \notag
\\
\hat{L}_{2} &=&\sigma \left\vert \phi \right\vert ^{2(\sigma -1)}\phi ^{2}.
\label{LL}
\end{eqnarray}%
The generic stationary states are stable if all the corresponding
eigenvalues $\omega $ have zero imaginary parts (the above-mentioned Townes'
solitons, that are actually degenerate states, feature a specific
subexponentially growing instability, which is accounted for by additional
zero eigenvalues).

\subsection{Exact solutions and the Vakhitov-Kolokolov stability criterion}

In the conservative version of the model, with $\gamma =0$, an exact real
solution to Eq. (\ref{ODE}), which may be considered as a soliton pinned to
the attractive defect, can be readily obtained:%
\begin{equation}
\phi _{\gamma =0}(x)=\left\{ \sqrt{k(\sigma +1)}\,\mathrm{sech}\left[ \sigma
\sqrt{2k}\,(|x|+\xi )\right] \right\} ^{1/\sigma },  \label{phi}
\end{equation}%
with real parameter $\xi >0$ determined by equation
\begin{equation}
\tanh \left( \sigma \sqrt{2k}\,\xi \right) =\frac{1}{\sqrt{2k}},
\label{tanh}
\end{equation}%
i.e.,%
\begin{equation}
\xi =\left( 2\sigma \sqrt{2k}\right) ^{-1}\ln \left( \frac{\sqrt{2k}+1}{%
\sqrt{2k}-1}\right) .  \label{xi}
\end{equation}%
The squared amplitude of the pinned soliton is%
\begin{equation}
A^{2}(\sigma,k)\equiv \phi _{\gamma =0}^{2}(x=0)=\left[ \left( 1+\sigma
\right) \left( k-1/2\right) \right] ^{1/\sigma }.  \label{A}
\end{equation}%
It follows from Eq. (\ref{A}) that the solutions exist with the propagation
constant exceeding a cutoff value,
\begin{equation}
k>k_{\mathrm{cutoff}}\equiv 1/2,  \label{min}
\end{equation}%
as the amplitude vanishes at $k\rightarrow 1/2$.

Further, a family of exact localized $\mathcal{PT}$-symmetric solutions to
Eq. (\ref{ODE}) can also be found in the presence of the gain-loss term,
i.e., for $\gamma >0$, similar to how it was done in Ref.~\cite{Thaw} for
the cubic nonlinearity ($\sigma =1$):%
\begin{eqnarray}
\phi _{\gamma >0}(x) &=&\phi _{\gamma =0}(x)\exp \left[ -i\,\mathrm{sgn}%
(x)\tan ^{-1}(\gamma )\right]   \notag \\
&\equiv &\phi _{\gamma =0}(x)\frac{1-i\gamma \,\mathrm{sgn}(x)}{\sqrt{%
1+\gamma ^{2}}},  \label{exact}
\end{eqnarray}%
where $\phi _{\gamma =0}(x)$ is the solution for $\gamma =0$, as given by
Eqs.~(\ref{phi}) and~(\ref{tanh}). The jump in the phase in expression (\ref%
{exact}) is produced by the singular term $-i\gamma \delta ^{\prime }(x)$ in
complex potential (\ref{delta}).

Note that the exact localized solution for the linearized system, with
nonlinear term $|\phi |^{2\sigma }\phi $ dropped in Eq. (\ref{ODE}), is
given by Eqs. (\ref{phi}) and (\ref{exact}) with $\xi \rightarrow \infty $.
This solution exists for the single value of the propagation constant, $k=1/2
$ [cf. Eq. (\ref{min})],%
\begin{equation}
\phi _{k=1/2,\gamma >0}^{\mathrm{(linear)}}(x)=\phi _{0}\exp \left[ -|x|-i\,%
\mathrm{sgn}(x)\tan ^{-1}(\gamma )\right] ,  \label{linear}
\end{equation}%
where $\phi _{0}$ is an arbitrary constant. In addition to this localized
mode existing at the single positive value of $k$, the linearized system
supports a continuous spectrum of delocalized modes with $k<0$:%
\begin{eqnarray}
\phi _{k<0,\gamma >0}^{\mathrm{(linear)}}(x) &\!\!=\!\!&\phi _{0}\sin \left[
\sqrt{-2k}\,|x|-\tan ^{-1}\left( \sqrt{-2k}\right) \right]   \notag \\
&&\quad \!\!\times \exp \!\left[ -i\,\mathrm{sgn}(x)\tan ^{-1}(\gamma )%
\right] .  \label{linear2}
\end{eqnarray}%
Solutions (\ref{linear}) and (\ref{linear2}) exist at all values of the
gain-loss strength $\gamma $, i.e., unlike the above-mentioned generic
situation, in the present linearized model the $\mathcal{PT}$ symmetry does
not suffer the breakdown with the increase of $\gamma $.

A fundamental characteristic of the family of localized states is its norm
(alias the integral power, in terms of spatial solitons in the optical
waveguide), which is given by the same expression for the solutions with $%
\gamma =0$ and $\gamma >0$:
\begin{gather}
N(\sigma ,k)\equiv \int_{-\infty }^{+\infty }\left\vert \phi _{\gamma
=0}(x)\right\vert ^{2}dx\qquad \qquad \qquad \qquad \qquad \quad \,\vspace{%
0.1in}  \notag \\
=\int_{-\infty }^{+\infty }\left\vert \phi _{\gamma >0}(x)\right\vert
^{2}dx\qquad \qquad \qquad \quad \,\,\,\vspace{0.1in}  \notag \\
=\sqrt{2}\,\frac{(\sigma +1)^{1/\sigma }}{\sigma }k^{1/\sigma -1/2}\qquad
\qquad \qquad   \notag \\
\qquad \qquad \times \!\int_{0}^{\infty }\!\!\left[ \!\mathrm{sech}%
\!\!\left( y\!+\!\frac{1}{2}\ln \!\left( \frac{\sqrt{2k}+1}{\sqrt{2k}-1}%
\right) \!\!\right) \!\!\right] ^{2/\sigma }\!\!\!dy,  \label{N}
\end{gather}%
%
%
%
%
%
%
%
%  where a new intrinsic parameter of the family is defined as%
%  \begin{equation}
%  q\equiv \sqrt{2k}.  \label{q}
%  \end{equation}%
% According to the existence region  the existence (\ref{min}),
In the limit of $k\rightarrow 1/2$, Eq.~(\ref{N}) demonstrates that the norm
vanishes as%
\begin{equation}
N(\sigma ,k)\approx \left[ \left( 1+\sigma \right) \left( \sqrt{2k}-1\right) %
\right] ^{1/\sigma }.  \label{q-1}
\end{equation}

According to the VK criterion~\cite{VKC,perturbation,Berge,Yang-book,Fibich}%
, the necessary stability condition in the conservative model, with $\gamma
=0$, is $\partial N/\partial k>0$. In particular, in the case of the usual
Kerr nonlinearity ($\sigma =1$), Eq. (\ref{N}) takes a very simple form~\cite%
{Thaw},
\begin{equation}
N(\sigma =1, k)=2\left( \sqrt{2k}-1\right),  \label{cubic}
\end{equation}%
which implies an evident result, that the pinned solitons are VK-stable
modes in the cubic medium. Similar to this result, the norm of the pinned
solitons diverges, at $k\rightarrow \infty $, for all subcritical
nonlinearities, with $\sigma <2$, as%
\begin{equation}
N(\sigma, k)\approx \sqrt{\frac{\pi }{2}}\frac{\left( \sigma +1\right)
^{1/\sigma }}{\sigma }\frac{\Gamma \left( 1/\sigma \right) }{\Gamma \left(
1/2+1/\sigma \right) }k^{1/\sigma -1/2},  \label{(k)}
\end{equation}%
where $\Gamma $ is the Gamma-function.

\begin{figure}[t]
\begin{center}
\vspace{0.05in} {\scalebox{0.40}[0.40]{\includegraphics{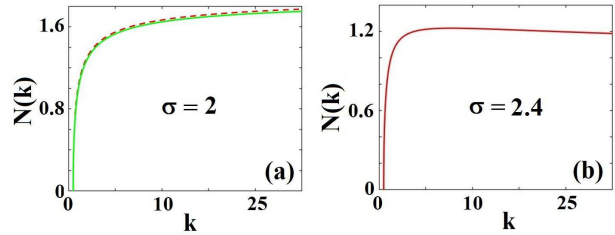}}}
\end{center}
\par
\vspace{-0.2in}
\caption{{\protect\small (Color online). (a) Norm $N(\protect\sigma ,k)$
versus parameter $k$ at $\protect\sigma =2$. The solid green line represents
the exact solution corresponding to Eq. (\protect\ref{sigma=2}), while the
dotted red line is produced by the numerical solution of Eq. (\protect\ref%
{phi}) with the delta-function regularized as per Eq. (\protect\ref{approx}%
). The application of the VK criterion to the $N(\protect\sigma =2,k)$\
dependence suggests that the entire family of the 1D Townes' solitons is
\emph{completely stabilized} by pinning to the attractive defect. (b) The
approximate dependence $N(\protect\alpha ,k)$ given by Eq. (\protect\ref%
{small-alpha}) with} $\protect\sigma =2.4$.}
\label{Nq}
\end{figure}

More interesting is the critical case of $\sigma =2$ (the quintic
nonlinearity), which was not considered previously in the combination with
the attractive defect. In this case, Eq.~(\ref{N}) yields
\begin{equation}
N(\sigma=2, k)=\sqrt{\frac{3}{2}}\left[ \pi -2\tan ^{-1}\left( \sqrt{\frac{%
\sqrt{2k}+1}{\sqrt{2k}-1}}\right) \right] ,  \label{sigma=2}
\end{equation}%
see Fig.~\ref{Nq}(a). The uniform medium [with $\varepsilon =0$, in terms of
Eq. (\ref{delta})] formally corresponds to $k\rightarrow \infty $, which
implies the degeneracy of the corresponding family of the 1D Townes'
solitons: their norm takes a single value,
\begin{equation}
N_{\mathrm{Townes}}=\sqrt{3/2}\left( \pi /2\right) \approx \allowbreak
1.\,\allowbreak 92,  \label{Townes}
\end{equation}%
which does not depend on the soliton's propagation constant. In terms of the
VK criterion, the constant norm formally corresponds to the neutral
stability, with $\partial N/\partial k=0$. However, it is well known that
the Townes' solitons are always unstable against the spontaneous onset of
the \textit{critical collapse}, although their instability is
subexponential. as mentioned above \cite{Fibich} (for this reason, it is not
detected by the VK criterion).

In the case of $\gamma =0$ and $\sigma =2$, Eq. (\ref{sigma=2}) clearly
demonstrates $\partial N/\partial k>0$ at all values of $k>\frac{1}{2}$ (see
Fig. \ref{Nq}), suggesting that the 1D Townes' solitons, which are
completely unstable in the free space, are \emph{completely stabilized} by
the attractive center, irrespective of its strength. This conjecture was
corroborated by the full stability analysis based on the numerical solution
of eigenvalue equation (\ref{lin}), as well as by direct simulations of the
perturbed evolution of the pinned solitons.

For the slightly supercritical case, with $2<\sigma\ll 3$, the $N(\sigma ,k)$
dependence derived from Eq. (\ref{N}) can be approximated by
\begin{equation}
N(\sigma ,k)\approx \sqrt{\frac{3}{2}}\,(2k)^{(2-\sigma)/4}\!\left[\pi
\!-\!2\tan ^{-1}\!\left( \!\sqrt{\frac{\sqrt{2k}+1}{\sqrt{2k}-1}}\!\right)
\!\!\right].  \label{small-alpha}
\end{equation}%
%
%
%
%
%
%
%
%A straightforward analysis of Eq. (\ref{small-alpha}) shows that $dN_{%
%\mathrm{small~}\alpha }(q)/dq=0$ at a large value of $q$, namely,%
%\begin{equation}
%q_{\mathrm{cr}}\approx 4/\left( \pi \alpha \right) \text{.}  \label{cr}
%\end{equation}
This dependence satisfies the VK criterion at
\begin{equation}
\frac{1}{2}<k<k_{\mathrm{cr}}\approx \frac{8}{\left( \pi (\sigma -2)\right)
^{2}},  \label{cr}
\end{equation}%
and does not satisfy it at $k>$ $k_{\mathrm{cr}}$. The approximate
dependence (\ref{small-alpha}) attains a maximum at $k=k_{\mathrm{cr}}$,
which, in the lowest approximation, turns out to be $N_{\max }=N_{\mathrm{%
Townes}}$, see Eq. (\ref{Townes}). Accordingly, the pinned solitons are
expected to be stable in this supercritical case (with $\gamma =0$, i.e., in
the absence of the gain and loss) in the region of $1/2<k<k_{\mathrm{cr}}$,
and to suffer the onset of the \textit{supercritical collapse} (spontaneous
blowup) at $k>k_{\mathrm{cr}}$. In particular, for $\sigma =2.4$ (not very
close to $\sigma =2$, but still relatively close), the dependence given by
Eq.~(\ref{small-alpha}) is plotted in Fig. \ref{Nq}(b), $\ $where $k_{%
\mathrm{cr}}\approx \allowbreak 7.75$, while approximation (\ref{cr}) yields
$k_{\mathrm{cr}}\approx 5.07$ for $\sigma =2.4$.

Another corollary of Eqs. (\ref{small-alpha}) and (\ref{cr}) is that
derivative $\partial N_{\max }/\partial \sigma $ slowly diverges at the
border of the supercritical case, i.e., at $\sigma \rightarrow 2$:%
\begin{equation}
\frac{\partial N_{\max }}{\partial \sigma }\approx -\sqrt{\frac{3}{2}}\frac{%
\pi }{4}\ln \left( \frac{1}{\sigma -2}\right) .  \label{deriv}
\end{equation}%
This result agrees with the fact that $N_{\max }=\infty $ at $\sigma <2$, as
it follows from Eq. (\ref{(k)}).

\begin{figure}[t]
\begin{center}
\vspace{0.05in} %\hspace{-0.05in}
{\scalebox{0.40}[0.40]{\includegraphics{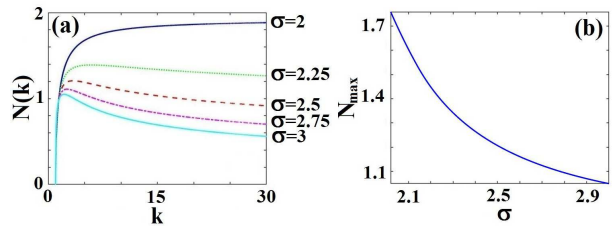}}}
\end{center}
\par
\vspace{-0.2in}
\caption{{\protect\small (Color online). (a) The dependence of norm $N(%
\protect\sigma ,k)$ on propagation constant }$k${\protect\small ,
numerically computed as per Eq.~(\protect\ref{N}), for the critical case,
with }${\protect\small \protect\sigma =2}${\protect\small , and the
supercritical one, with }${\protect\small \protect\sigma =}${\protect\small $%
\ 2.25,\,2.5,\,2.75,\,3$. (b) The maximum norm of the pinned solitons,
attained at the stability-boundary, }$k=k_{\mathrm{cr}}${\protect\small , at
which $\partial N/\partial k=0$, versus $\protect\sigma $\thinspace\ (in the
supercritical interval, $2<\protect\sigma \leq 3$). In the case of }$%
{\protect\small \protect\gamma =0}${\protect\small , the pinned solitons are
stable in the region of }${\protect\small 1/2<k<k}_{\mathrm{cr}}$%
{\protect\small , which corresponds to }${\protect\small 0<N<N}_{\max }$%
{\protect\small .}}
\label{Nq_numerical}
\end{figure}

Figure~\ref{Nq_numerical} displays the dependence $N(\sigma ,k)$,
numerically calculated as per Eq. (\ref{N}), for different critical and
supercritical values of the nonlinearity power, \textit{viz}., $\sigma
=2,\,2.25,\,2.5,\,2.75,\,3$. While, as said above, in the critical case of $%
\sigma =2$ condition $\partial N(\sigma ,k)/\partial k>0$ holds at all $k>%
\frac{1}{2}$, at each supercritical value, $\sigma >2$, the $N(\sigma ,k)$
dependence indeed features the critical point, $k_{\mathrm{cr}}$, at which
the slope, $\partial N/\partial k$, changes its sign, and the norm attains
its largest value, $N_{\max }=N\left( k_{\mathrm{cr}}\right) $ [note that,
as follows from Eq. (\ref{N}), $N\left( \sigma ,k\rightarrow \infty \right)
\sim (2k)^{\left(1/\sigma-1/2 \right) }\rightarrow 0$, for any $\sigma >2$].
The prediction of the VK criterion, that, in the case of $\gamma =0$, the
pinned solitons are stable in the region of $1/2<k<k_{\mathrm{cr}}$, i.e., $%
0<N<N_{\max }$, is corroborated by the calculation of the stability
eigenvalues via Eq. (\ref{lin}), as well as by direct simulations of
perturbed evolutions of the solitons. In particular, the VK-unstable
solitons, existing at $k>k_{\mathrm{cr}}$, are indeed destructed by the
spontaneous blowup, see Fig. \ref{Evolution}(a) below. Thus, the attractive
defect provides for the partial stabilization of the solitons in the case of
the supercritical nonlinearity, for all values of $\sigma >2$, while all
solitons are strongly unstable in this case in the absence of the defect.
Naturally, Fig.~\ref{Nq_numerical}(a) demonstrates that the stability region
(in both forms of $1/2<k<k_{\mathrm{cr}}$ and $0<N<N_{\max }$), maintained
by the interplay of the supercritical self-attractive nonlinearity and
attraction center, shrinks with the increase of the nonlinearity power, $%
\sigma $. Nevertheless, the stability region persists even at large values
of $\sigma $. Indeed, if $\sigma $ is treated as a large parameter, while $k-%
\frac{1}{2}$ as a small one, Eq. (\ref{N}) amounts to
\begin{equation}
N(\sigma\!\gg 1, k)\approx (2k)^{-1/2}(\sqrt{2k}-1)^{1/\sigma }
\label{large-sigma}
\end{equation}%
[cf. Eq. (\ref{q-1})], which yields%
\begin{equation}
k_{\mathrm{cr}}-\frac{1}{2}\approx 1/\sigma ,\quad ~\lim_{\sigma \rightarrow
\infty }N_{\max }=1.  \label{lim}
\end{equation}

\begin{figure}[t]
\begin{center}
\vspace{0.05in} \hspace{-0.05in}{\scalebox{0.45}[0.56]{%
\includegraphics{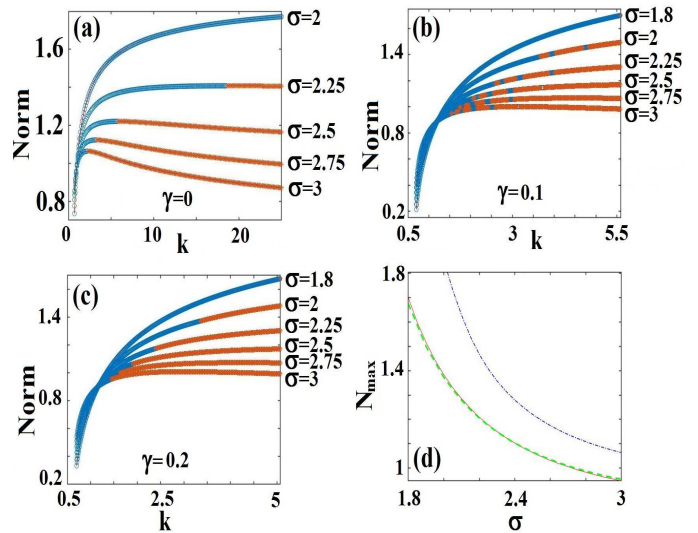}}}
\end{center}
\par
\vspace{-0.25in}
\caption{{\protect\small (Color online). The norm of stable solitons (blue
circles) and unstable solitons (red stars) produced by the numerical
solution of Eq.~(\protect\ref{phi}) with the regularized delta-function (%
\protect\ref{approx}), for different values of the nonlinearity degree $%
\protect\sigma $ and gain-loss coefficient $\protect\gamma $: (a) }$%
{\protect\small \protect\gamma =0}${\protect\small , (b) }${\protect\small
\protect\gamma =0.1}${\protect\small , (c) }${\protect\small \protect\gamma %
=0.2}${\protect\small . The stability is identified through the numerical
solution of eigenvalue problem (\protect\ref{lin}), using the same
regularization. Note the difference in the scales of the horizontal axes
between panels (a) and (b,c). The top and bottom curves in panel (d) display
stability boundaries in the plane of ($\protect\sigma $, }${\protect\small N}
${\protect\small ) for }${\protect\small \protect\gamma =0}$ (dash-dotted
blue line){\protect\small ,} ${\protect\small 0.1}$ (solid red line)%
{\protect\small , and }${\protect\small 0.2}$ (dashed green line)%
{\protect\small \ (the boundaries for }${\protect\small \protect\gamma =0.1}$%
{\protect\small \ and }${\protect\small \protect\gamma =0.2}${\protect\small %
\ are practically identical).}}
\label{dGamma}
\end{figure}

The limit value of $N_{\max }=1$, given by Eq. (\ref{lim}), agrees with the
numerical results displayed in Figs. \ref{Nq_numerical}(b) and \ref{dGamma}%
(d).

Finally, to better understand the physical meaning of the $\mathcal{PT}$%
-symmetric stationary state, $\phi _{\gamma >0}(x)$, given by Eq.~(\ref%
{exact}), it is relevant look at its power flux (the Poynting vector), $%
S(x)\!=\mathrm{Im}\left( \phi ^{\ast }(x)\frac{d}{dx}\phi (x)\right) $. The
substitution of Eqs. (\ref{ODE}) and (\ref{exact}) yields a singular
expression,%
\begin{equation}
S(x)=-2\gamma \phi _{\gamma =0}^{2}(x)\delta (x),  \label{S}
\end{equation}%
A regular characteristic of the transport in the $\mathcal{PT}$-symmetric
pattern is provided by the globally normalized flux,
\begin{equation}
\begin{array}{rl}
\hat{S}\equiv & \displaystyle N^{-1}(\sigma,k)\int_{-\infty }^{+\infty
}S(x)dx \vspace{0.1in} \\
= & -2\gamma A^{2}(\sigma,k)/N(\sigma,k)<0,%
\end{array}
\label{flux}
\end{equation}%
where $A(\sigma,k)$ is the amplitude of the pinned soliton, given by Eq. (%
\ref{A}). An essential feature of this expression is its sign, which, in
comparison with Eq. (\ref{NLS}), clearly demonstrates that the power flows,
as it should, from the source (gain) towards the sink (loss).%
%\begin{equation}
%\begin{array}{rl}
%S(x)\!= & \!\!\dfrac{1}{2}\left[ \phi _{\gamma >0}(x)\frac{d}{dx}\phi
%_{\gamma >0}^{\ast }(x)-\phi _{\gamma >0}^{\ast }\frac{d}{dx}\phi _{\gamma
%>0,x}(x)\right] ,\vspace{0.1in} \\
%\!= & \!\!-2A^{2}\tan ^{-1}(\gamma )\vspace{0.1in} \\
%\!= & \!\!-2\left[ \left( 1+\sigma \right) \left( k-1/2\right) \right]
%^{1/\sigma }\tan ^{-1}(\gamma )\vspace{0.1in} \\
%\!< & \!\!0,\quad \mathrm{for}\,\,k>1/2,%
%\end{array}%
%\end{equation}%
%which implies that the power always flows from the gain toward the loss.

\subsection{Exact solutions with the self-defocusing nonlinearity}

It is relevant to briefly consider the model with the self-repulsive uniform
nonlinearity, i.e., with Eq. (\ref{NLS}) replaced by%
\begin{equation}
i\psi _{z}=-\frac{1}{2}\psi _{xx}+\left\vert \psi \right\vert ^{2\sigma
}\psi -[\varepsilon \delta (x)+i\gamma \delta ^{\prime }(x)]\psi.
\label{defocusing}
\end{equation}%
In this case, the exact solution for $\gamma =0$ is readily found as%
\begin{equation}
\phi _{\gamma =0}^{\mathrm{(defoc)}}(x)=\left\{ \sqrt{k(\sigma +1)}/\sinh %
\left[ \sigma \sqrt{2k}\,(|x|+\xi )\right] \right\} ^{1/\sigma },
\label{defoc}
\end{equation}%
with%
\begin{equation}
\xi =\frac{1}{2\sigma \sqrt{2k}}\ln \left( \frac{1+\sqrt{2k}}{1-\sqrt{2k}}%
\right) ,  \label{xi-defoc}
\end{equation}%
and the squared amplitude%
\begin{equation}
B^{2}(\sigma,k)\equiv \left[ \phi _{\gamma =0}^{\mathrm{(defoc)}}(x=0)\right]
^{2}=\left[ \left( 1+\sigma \right) \left( 1/2-k\right) \right] ^{1/\sigma }.
\label{A-defoc}
\end{equation}%
As it follows from Eq. (\ref{A-defoc}), the existence region for the
localized modes pinned to the attractive defect embedded in the defocusing
medium is $k<k_{\mathrm{cutoff}}\equiv 1/2$, which is exactly opposite to
that in the case of self-focusing, cf. Eq. (\ref{min}).

In the presence of the gain-loss dipole, which introduces the $\mathcal{PT}$
symmetry, i.e., $\gamma >0$, the exact solution is generated from the one
for $\gamma =0$ by the same relation (\ref{exact}) which is relevant in the
case of self-focusing. As concerns the norm, it takes a simple form in the
cubic defocusing medium ($\sigma =1$),
\begin{equation}
N^{\mathrm{(defoc)}}(\sigma =1, k)=2\left( 1-\sqrt{2k}\right)
\label{sigma=1}
\end{equation}%
[cf. Eq. (\ref{cubic})]. On the other hand, in the limit case of large $%
\sigma $, the result is
\begin{equation}
N^{\mathrm{(defoc)}}(\sigma\!\gg 1, k)\approx \left( 2k\right) ^{-1/2}\left(
1-\sqrt{2k}\right) ^{1/\sigma },  \label{sigma>>1}
\end{equation}%
cf. Eq. (\ref{large-sigma}). It can be checked that, as well as it is
obvious in Eqs. (\ref{sigma=1}) and (\ref{sigma>>1}), the $N(\sigma, k)$
dependence corresponding to the defocusing nonlinearity satisfies the
above-mentioned anti-VK criterion, $\partial N/\partial k<0$ \cite{anti-VK},
at all values of $\sigma $. Accordingly, all the pinned modes are stable,
both at $\gamma =0$ and $\gamma >0$, similar to the case of the
self-defocusing cubic nonlinearity \cite{Thaw}.

\section{Numerical solutions and their stability}

\begin{figure}[t]
\begin{center}
\hspace{-0.05in}{\scalebox{0.70}[0.70]{\includegraphics{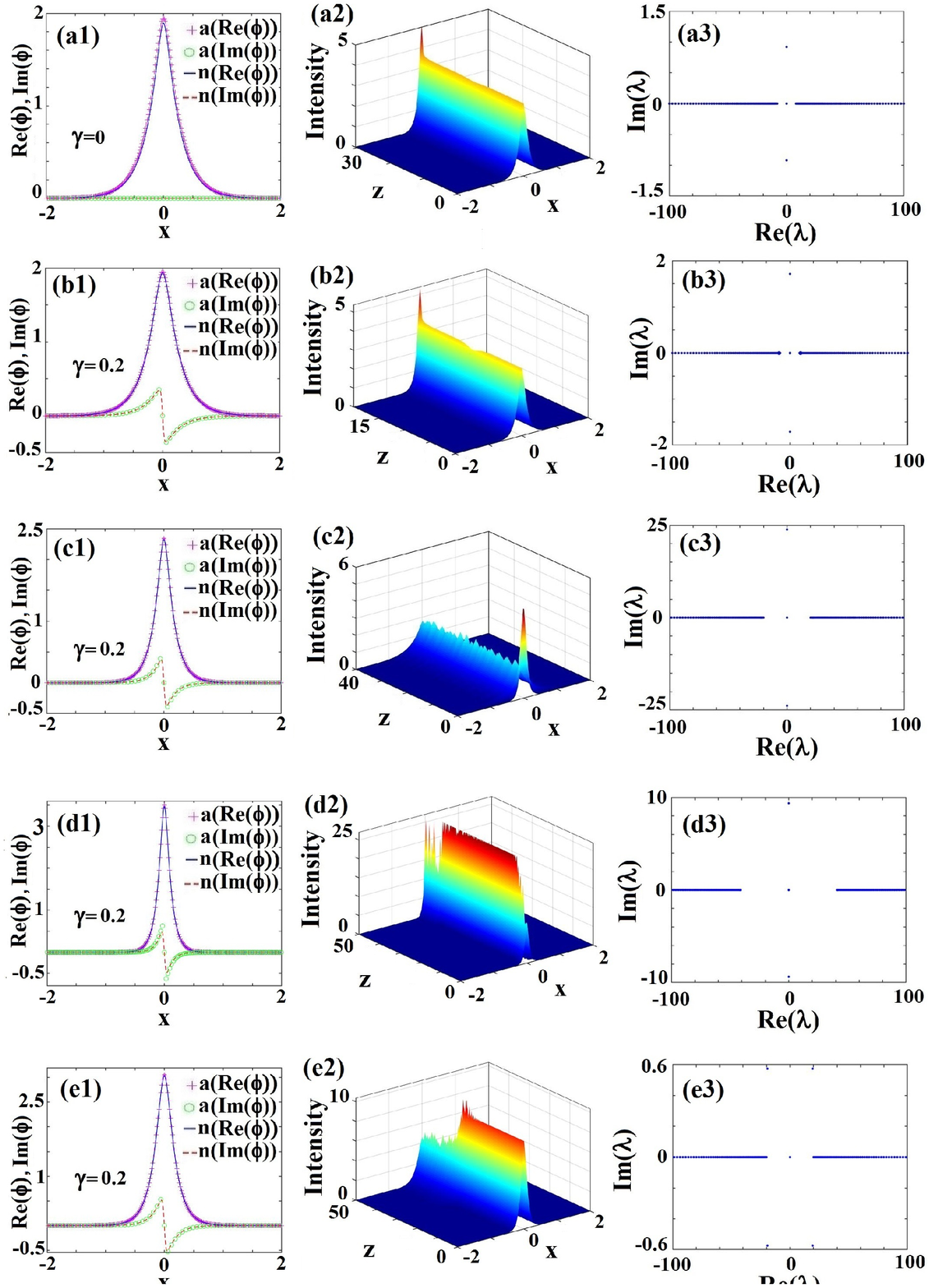}}}
\end{center}
\par
\vspace{-0.2in}
\caption{{\protect\small (Color online). (a1), (b1), (c1), (d1), and (e1):
Comparison between typical examples of unstable analytical solutions,
labeled \textquotedblleft a" (magenta pluses and green circles display their
real and imaginary parts, respectively), given by Eqs.~(\protect\ref{phi}), (%
\protect\ref{xi}), and (\protect\ref{exact}), and their counterparts
obtained from the numerical solution of Eq. (\protect\ref{ODE}), subject to
regularization~(\protect\ref{approx}), labeled \textquotedblleft n", whose
real and imaginary parts are shown by blue solid and red dashed lines,
respectively. A small difference between the analytically predicted and
numerically found imaginary parts of the solutions is a result of the
regularization, as the analytical version is discontinuous, see Eq. (\protect
\ref{exact}). The simulated evolution of the numerical solutions from (a1),
(b1), (c1), (d1), and (e1) is displayed, severally, in panels (a2), (b2),
(c2), (d2), and (e2), and their stability spectra are depicted in (a3),
(b3), (c3), (d3), and (e3), respectively. The parameters are: $\protect%
\gamma =0,\,\protect\sigma =2.5,\,k=8$ in (a1, a2, a3), $\protect\gamma %
=0.2,\,\protect\sigma =2.5,\,k=8$ in (b1, b2, b3), $\protect\gamma =0.2,\,%
\protect\sigma =2.5,\,k=20$ in (c1, c2, c3), $\protect\gamma =0.2,\,\protect%
\sigma =2,\,k=50$ in (d1, d2, d3), and $\protect\gamma =0.2,\,\protect\sigma %
=1.8,\,k=20$ in (e1, e2, e3).}}
\label{Evolution}
\end{figure}

\subsection{Stationary states and small perturbations}

While the model with the ideal delta-functional defect admits exact
solutions for all values of $\sigma $ and $\gamma $, as given by Eqs. (\ref%
{phi}), (\ref{xi}), and (\ref{exact}), their stability in the $\mathcal{PT}$%
-symmetric model with $\gamma >0$ is a nontrivial problem which should be
addressed by means of numerical methods. For this purpose, the ideal
delta-function is replaced by the well-known Gaussian regularization,
\begin{equation}
\delta (x)\rightarrow \widetilde{\delta }(x)=\frac{1}{\sqrt{\pi }a}\exp
\left( -\frac{x^{2}}{a^{2}}\right) ,\vspace{0.1in}  \label{approx}
\end{equation}%
with width $a$ much smaller than a characteristic size of the pinned mode.
This condition is secured by fixing $a=0.03$, which is adopted here. It was
checked that taking smaller $a$ does not affect the results in any
conspicuous way.

Stationary equation (\ref{phi}) with $\delta (x)$ and $\delta ^{\prime }(x)$
replaced by $\widetilde{\delta }(x)$ and $\widetilde{\delta }^{\prime }(x)$
can be efficiently solved by means of the Petviashvili iteration method~\cite%
{Pmethod}, using the above-mentioned analytical solution, valid in the limit
of the ideal delta-function, as the initial guess. Then, Eqs. (\ref{lin})
and (\ref{LL}), regularized as per substitution (\ref{approx}), are
numerically solved to predict the stability boundary. Finally, perturbed
evolution of the pinned modes was simulated by means of fourth-order
Runge-Kutta time-stepping scheme. A known necessary and sufficient
condition~for the stability of the direct simulations amounts to an
inequality imposed on the time and space steps, $\Delta t/\left( \Delta
x\right) ^{2}\leq 2\sqrt{2}/\pi ^{2}$ \cite{Yang-book}. As a result, it is
concluded that the real numerically generated solutions for $\gamma =0$, as
well as real parts of the complex solutions for $\gamma >0$, are very close
to the analytically predicted counterparts [see Figs. \ref{Evolution}%
(a1,b1,c1,d1,e1)]. A small difference of the imaginary parts in the latter
case (hence, of the phase structure of the complex solitons) from the
analytical form given by Eq. (\ref{exact}) is explained by the fact that the
numerical solution cannot exactly follow the discontinuous shape of the
solution produced by Eq. (\ref{exact}).

Figure~\ref{dGamma}(a) summarizes results for the stability of the pinned
solitons, produced by the numerical solution of Eqs. (\ref{ODE}) and (\ref%
{lin}) subject to linearization (\ref{approx}), for the critical and
supercritical values of the nonlinearity power, $\sigma =\,2$ and $%
\,2.25,\,2.5,\,2.75,\,3$, in the absence of the gain and loss ($\gamma =0$).
It is seen that the stability precisely follows the VK criterion, and, in
the exact agreement with the analytical prediction, the solitons are
completely stable for $\sigma =2$, while the stability region shrinks with
the increase of $\sigma $ in the supercritical case. The respective
existence and stability boundary, $N_{\max }(\sigma )$, which is shown by
the top curve in Fig. \ref{dGamma}(d), is identical to its analytically
predicted counterpart, cf. Fig. \ref{Nq_numerical}(b).

As it may be expected, the stability pattern becomes more complex in the
presence of the gain-and-loss term, i.e., at $\gamma >0$, see Figs. \ref%
{dGamma}(b) and (c) for $\gamma =0.1$ and $0.2$. It is observed that the VK
criterion is no longer sufficient for the stability, in this case. For the
relatively large value $\gamma =0.2$, Fig. \ref{dGamma}(c) demonstrates
strong shrinkage of the stability region, in comparison with the case of $%
\gamma =0$; in particular, the critical nonlinearity, with $\sigma =2$,
keeps the pinned solitons stable only at $N<N_{\max }\approx 1.32$, while,
at $\gamma =0$, the same nonlinearity maintains the stability in the entire
existence region of the solitons, up to $N=N_{\mathrm{Townes}}$, see Eq.~(%
\ref{Townes}). The numerically found stability boundaries for $\gamma =0.1$
and $0.2$, in the form of $N_{\max }(\sigma )$, are displayed by two bottom
curves\ (which are virtually identical to each other) in Fig.~\ref{dGamma}%
(d).

Figure \ref{dGamma}(b) features an intricate structure of the stability
pattern, with\emph{\ multiple stability intervals}, at smaller values of the
gain-loss coefficient, such as $\gamma =0.1$. As seen in Fig. \ref{dGamma}%
(d), the lowest stability boundary for $\gamma =0.1$ is virtually identical
to one for $\gamma =0.2$. However, the actual number of such boundaries
observed in \ref{dGamma}(b) is, at least, five. Probably, many more
boundaries may be revealed by extremely precise numerical data, and it seems
plausible that the exact structure of the stability islands, embedded in the
instability area, may be \textit{fractal}, as suggested, in particular, by
the fractal alternation of regions of elastic and inelastic collisions
between solitons in some nonintegrable conservative models \cite%
{Campbell1,Campbell2,Yang-collision}. The fractal structure, if any, may
strongly depend in values of $\gamma $ and $\sigma $. A rigorous analysis of
this challenging problem is beyond the scope of the present work.

In the presence of the gain and loss, the instability occurs even in the
subcritical case, i.e., at $\sigma <2$. While in Figs.~\ref{dGamma}(b, c)
the subcritical branches, corresponding to $\sigma =1.8$, are completely
stable in the displayed range of $k$, they develop weak instability at
essentially larger values of $k$, where the VK derivative, $\partial
N/\partial k\sim k^{1/\sigma -3/2}$ [see Eq. (\ref{(k)})] is very small,
hence the the effect of the subcriticality is insufficient to suppress the
instability induced by the possibility of spontaneous breaking of the $%
\mathcal{PT}$ symmetry. In particular, for $\sigma =1.8$ and $\gamma =0.2$,
the pinned soliton is unstable at $k=20$, see Figs. \ref{Evolution}(e2,e3)
below.

\subsection{Direct simulations}

The predictions for the (in)stability of the pinned solitons, produced by
the application of the VK (or anti-VK) criterion to the analytical solutions
in the case of $\gamma =0$, and by the numerical solution of Eq. (\ref{lin}%
), subject to regularization (\ref{approx}), for the numerically constructed
solutions in the case of $\gamma \geq 0$, were verified by direct
simulations of perturbed evolutions of the pinned solitons. It has been
concluded that all the modes which were predicted to be stable are stable
indeed (not shown here in detail, as simulations of the stable evolution do
not reveal new features of the dynamics). The predicted instability is also
corroborated by the direct simulations, as shown in Fig.~\ref{Evolution}. In
particular, in the case of $\gamma =0$ the instability, displayed in panel
(a2), leads to sudden blowup of the solution, which is a manifestation of
the supercritical collapse. The same happens too, but faster, as a result of
the interplay of the supercritical self-attraction and gain-loss ($\gamma
=0.2$) term, see panel (b2). Panel (c2) demonstrates that, in the case of
large $k$, when the instability related to the supercriticality is weak, as
the derivative, $\left\vert \partial N/\partial k\right\vert $, which
determines the VK criterion, is small, see Figs. \ref{Nq}(b) and \ref%
{Nq_numerical}(a), the interplay of the weak instability and gain-loss terms
leads not to a blowup, but to the formation of a robust localized breather
with an amplitude which is essentially smaller than in the original unstable
soliton.

Recall that, in the absence of the gain-loss term, all the pinned solitons
maintained by the critical nonlinearity, with $\sigma =2$, are stable. The
addition of the sufficiently strong gain and loss [in particular, with $%
\gamma =0.2$, as shown in Fig. \ref{Evolution}(d2)] may destabilize the
solitons in this case. It is observed in the figure that the initial stage
of the instability-driven evolution is chaotic, eventually relaxing to a
robust breather.

Finally, as it was mentioned above, the presence of the gain-loss term may
induce instability in the subcritical case, $\sigma <2$. Figure \ref%
{Evolution}(e2) demonstrates an example of this instability for $\sigma =1.8$
and $\gamma =0.2$, provided that the soliton was created with a large value
of the propagation constant (in particular, $k=20$ in this figure), where,
as said above, the stabilizing effect of the subcritical instability is very
weak. In this case too, the collapse does not occur (as there is nothing to
drive it in the subcritical case), the instability leading to spontaneous
transformation of the unstable soliton into a robust breather with an
essentially smaller amplitude.

As seen from Fig. \ref{dGamma}, the stability region of the $\mathcal{PT}$%
-symmetric solitons shrinks with the increase of both the nonlinearity
power, $\sigma $, and gain-loss coefficient, $\gamma $. Systematically
collecting numerical data, we conclude that, at a fixed value of $\sigma $,
all the solitons become unstable at $\gamma >\gamma _{\max }(\sigma )$, and,
at a fixed value of $\gamma $, the same happens at $\sigma >\sigma _{\max
}(\gamma )$. %These results are summarized by the stability boundary in the
%plane of $\left( \sigma ,\gamma \right) $, which is plotted in Fig. \ref%
%{sigma_Gamma}.
As concerns value $N_{\max }$ of the norm at which the stability ends, it
essentially depends on $\sigma $ and very weakly depends on $\gamma $, as
shown by the stability boundaries in the $\left( \sigma ,N\right) $ plane
for $\gamma =0.1$ and $0.2$ in Fig. \ref{dGamma}(d), which are virtually
identical for both these values of $\gamma $. For a fixed value of the norm,
$N<N_{\max }$, the dependence of the stability border of $\gamma $ is very
weak too.

%%%%%%%%%%%%%%%%%%%%%%%%%%%%%%%%%%%%%%%%%%%%%%%%%%%%%
%\begin{figure}[!t]
%\begin{center}
%\vspace{0.05in} \hspace{-0.05in}{\scalebox{0.7}[0.5]{%
%\includegraphics{Norm_dSigma}}}
%\end{center}
%\par
%\vspace{-0.55in}
%\caption{{\protect\small (color online). Norm versus the $\mathcal{PT}$
%gain-loss coefficient $\protect\gamma$ in Eq.~(\protect\ref{appro-ODE}) for
%fixed $\protect\sigma$. (a) $\protect\sigma$=1.9. (b) $\protect\sigma$=2.0.
%(c) $\protect\sigma$=2.25. (d) $\protect\sigma$=2.50. (e) $\protect\sigma$%
%=2.75. (f) $\protect\sigma$=3.0.}}
%\label{Norm_dSigma}
%\end{figure}
%%%%%%%%%%%%%%%%%%%%%%%%%%%%%%%%%%%%%%%%%%%%%%%%%%%%
%\begin{table}[tbp]
%\caption{The stability for $\protect\sigma$ and critical points $\protect%
%\gamma$. }
%\label{table1}%
%\begin{tabular}{ccccccc}
%\hline\hline
%&  &  &  &  &  &  \\[-2.0ex]
%$\sigma$ & 1.9 & 2.0 & 2.25 & 2.5 & 2.75 & 3 \\[1.0ex] \hline
%&  &  &  &  &  &  \\[-2.0ex]
%$\gamma$ & ~0.107 & ~0.021 & ~0.016 & ~0.013 & ~0.012 & ~0.011 \\%
%[1.0ex] \hline\hline
%\end{tabular}%
%\end{table}

\section{Conclusions and discussions}

The objective of this work is to introduce settings in which intrinsically
unstable solitons in 1D media modelled by the NLSE with the critical ($%
\sigma =2$) and supercritical ($\sigma >2$) self-attractive nonlinearity may
be stabilized by pinning to an attractive defect, including its $\mathcal{PT}
$-symmetric version with strength $\gamma $ of the gain-loss component. The
settings represent a planar nonlinear-optical waveguide, or a
super-Tonks-Girardeau gas. A remarkable fact is that full families of the
pinned solitons can be found in the analytical form for the delta-functional
defect, and, in the absence of the gain-loss term, their stability too
admits analytical investigation, by means of the VK criterion. In
particular, a stability interval exists for arbitrarily high values of the
nonlinearity power $\sigma $. In the presence of the gain and loss, the
stability of the $\mathcal{PT}$-symmetric soliton families and the evolution
of unstable solitons were investigated by means of numerical methods,
revealing a nontrivial stability area in the $\left( \sigma ,\gamma \right) $
plane. For relatively small values of $\gamma $, the structure of the
stability area is intricate, featuring multiple stability boundaries (which
may presumably form a fractal). If the instability driven\ by the
supercritical nonlinearity is strong enough, unstable solitons are destroyed
by the collapse, both in the absence and presence of the gain-loss term. On
the other hand, if the supercritical instability is weak, it does not lead,
in the combination with the balanced gain and loss ($\gamma >0$), to the
collapse; instead, it spontaneously replaces unstable solitons by robust
localized breathers. The same happens to those solitons which are unstable
under the action of the critical nonlinearity combined with the gain-loss
term. Instability is also possible if the gain and loss are added to the
subcritical instability, which is close to the critical case. In the latter
case, breathers emerge as well. The model with the self-repulsion was
briefly considered too, with the conclusion that the localized modes pinned
to the attractive defect are completely stable in that case.

A challenging direction for the extension of the present analysis is to
develop it in a two-dimensional model, which may also be realized in
nonlinear optics, using a bulk waveguide, with the local defect represented
by an elongated core embedded in the medium.

\vspace{0.1in} \acknowledgments %The authors would like to thank the referees
%for the valuable suggestions.
The present work was supported in the
framework of NSFC (China) under grants Nos. 11571346, 11731014, and
61621003, Interdisciplinary Innovation Team of Chinese Academy of Sciences,
and the Chinese Academy of Sciences President's International Initiative
(PIFI).

\end{document}